# Energy deposition in air by moderately focused femtosecond laser filaments

Yu.E. Geints[1], I.Yu. Geints[2], Ya.V. Grudzin[3], A.V. Koribut[3],
D.V. Pushkarev[3], G.E. Rizaev[3], L.V. Seleznev[3]

[1] *V. E. Zuev Institute of Atmospheric Optics SB RAS, 1, Acad. Zuev Square, Tomsk 634055, Russia*
[2] *Faculty of Physics, Lomonosov Moscow State University, Leninskie Gory, Moscow 119991, Russia*
[3] *P. N. Lebedev Physical Institute of the Russian Academy of Sciences, 53 Leninskii pr., Moscow 119991, Russia*

Abstract

Filamentation of high-power femtosecond laser pulses in air is accompanied by a fairly strong release of optical energy into the propagation medium due to laser-induced ionization of air molecules and production of an underdense plasma of charged species. We present the results of our laboratory experiments and numerical simulations aimed to the estimation of energy deposition amount by laser filament upon propagation in air depending on the conditions of spatial focusing, pulse energy, and radiation wavelength. For the first time to our knowledge, our study reveals a more than 50% decrease in the filament energy deposited in air in the range of moderate numerical aperture values, approximately from 0.003 to 0.007, at the carrier wavelengths of 740 nm and 470 nm. We attribute such a considerable reduction in the laser pulse energy release for femtosecond plasma to the competing effects of Kerr self-focusing and geometric divergence of focused laser pulse.

## 1. Introduction

Self-focusing and filamentation of high-power femtosecond laser pulses in air is actively studied since the mid-1990s [1-4]. The growing attention to this problem is due to the fact that laser filamentation is a prominent manifestation of the nonlinear optics and possesses great practical value for various applications, such as higher harmonic generation [5], atmospheric LiDAR applications [6-8], optical energy delivery to the remote targets [9, 10], remote fuel ignition [11, 12], etc. Self-focusing of optical radiation is realized due to the optical Kerr effect at the peak pulse power exceeding certain critical value. In this case, a self-induced convex lens is formed in medium which permanently focuses the optical wave as it propagates. As a result of such nonlinear self-focusing, a narrow high-intensity light beamlet, or a group of beamlets, usually called laser filaments, is formed within the mother beam. During the filamentation, pulse spectral composition is significantly enriched as a result of strong self-phase modulation in the nonlinear medium. This is manifested in the formation of broad supercontinuum spectral bands covering in some cases several octaves of initial pulse spectrum [5]. High intensity optical field inside the filaments causes multiphoton and tunnel ionization of medium molecules, which produces plasma channels accompanying laser radiation with a characteristic density of free electrons $10^{15}$ - $10^{18}$ cm$^{-3}$ [1].

Recently, plasma filaments were used to create airborne optical waveguides formed by rarefied regions of air emerging after the passage of a laser filament and persisting for tens of milliseconds depending on the gas-dynamic state of the surrounding air [13]. The study of such laser-induced air thermodynamics seems important because it paves the way to the development of remotely generated long-lived virtual waveguiding optical structures, which in the long term can be used for optical communication [14-16]. The efficiency of these gas-dynamic structures essentially depends on the amplitude of laser-induced effects, namely, on the pulse energy locally deposited in the gas medium by the filament, because the magnitude and spatial distribution of energy release determine the depth and axial extent of air refractive index perturbations. Therefore, it becomes critical to gain control on the amount of this energy release in the required region of pulse propagation path.

In previous studies it was found that the amount of pulse energy deposition in medium depends not only on optical energy, but also on pulse temporal duration [14, 17, 18], as well as on the conditions of external focusing [14, 19, 20]. In the latter case, it was shown that in general, the laser energy deposition is larger as the pulse is focused tighter (at fixed pulse energy). However, at very strong pulse focusing (relative aperture *f*/3) when the propagation regime close to the optical breakdown is realized, the value of the deposited optical energy on the contrary is reduced which was attributed to strong plasma defocusing of laser pulse in the nonlinear focal area [Point2016] due to the formation of a "superfilament" structure [21].

The present work is aimed at the determination of energy deposition value in air by the focused filamentation of high-power femtosecond laser pulses. Unlike previously published works, in our study we consider the range of moderate pulse focusing with the relative aperture less than *f*/40 and two spectral bands of laser radiation with the carrier at 740 nm and 470 nm. By means of the numerical simulations, we analyze the physical reasons leading to the dependences observed in the experiment. It turns out that even under these conditions, one can distinguish a certain region of focal distances, when laser pulse energy losses for femtosecond plasma generation are significantly reduced and optical energy deposition in air considerably drops. We attribute this drop with the competition of nonlinear self-focusing and linear geometric divergence of initially focused laser radiation.

## 2. Results and discussion

In our experiments, the optical radiation delivered from two separate titanium-sapphire laser oscillators is used, generating pulses with central wavelengths of 740 nm and 470 nm (second harmonic). Laser pulse energy can vary from 2 to 4 mJ using a diffraction attenuator. Pulse duration for both wavelengths is $t_p = 100$ fs. External focusing of laser radiation is performed by a spherical mirror and silica lenses with different focal distances $f$. Usually, the filamentation of a laser pulse occurred near the geometrical focus with the formation of an extended filamentary plasma channel. The side luminescence of this plasma channel is projected with a collecting lens onto a CCD matrix, and then the images are digitized and processed on a computer.

For the estimation of the amount of pulse energy deposition into the propagation medium, we measure the filament plasma luminescence. As demonstrated in [22], for plasma diagnostics the luminescence intensity $F$ of the molecular nitrogen cation can be used. In this case, the total number of free electrons $N_e$ in the resulting plasma is proportional to the integral of $F$ over the entire region **R** occupied by the plasma region:

$$N_e \propto B = \int_R F(\mathbf{r}) d\mathbf{r} \qquad (1)$$

In the experiments, the parameter $B$ is calculated by integrating the CCD energy response $F(\mathbf{r})$ over the entire image area of plasma channel after the noise subtraction procedure, and then the resulting values are averaged over several hundred pulses.

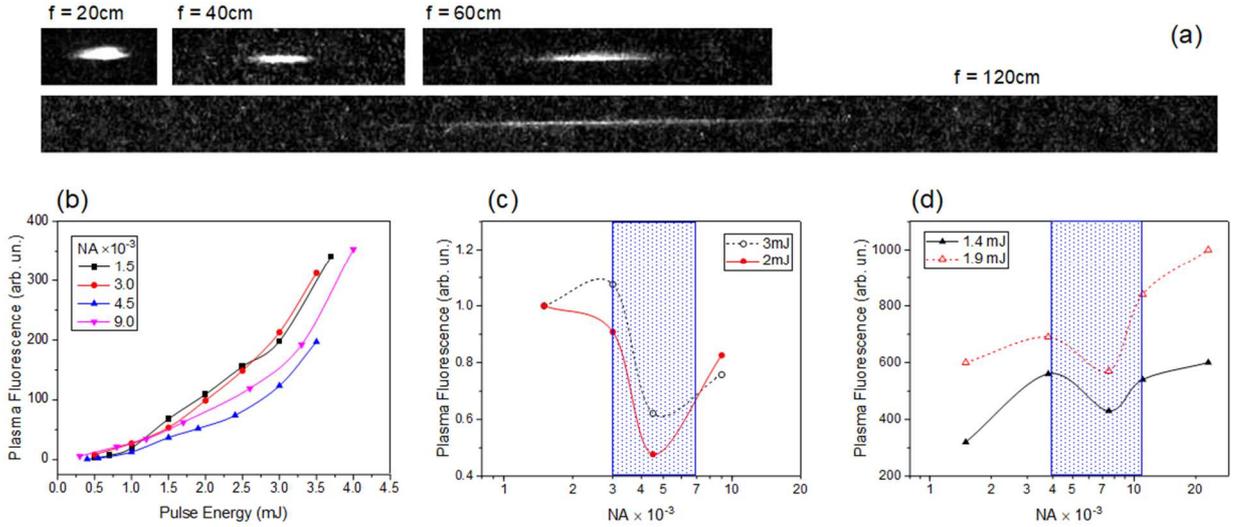

Fig. 1. Experiments on the plasma fluorescence diagnostics during femtosecond laser pulse filamentation. (a) CCD images of plasma channels (side view) at different pulse focusing. (b-d) Integral fluorescence $B$ of plasma filament produced by (b, c) 740 nm and (d) 470 nm laser pulses upon varying pulse energy and external focusing. NA range with apparent fluorescence changes is highlighted.

Representative images of plasma fluorescence in optical filament created by the focused femtosecond laser pulses in air are shown in Fig. 1(a) for different focusing conditions. In the following, it is practical to use the dimensionless parameter of the numerical aperture, NA = $D / 2f$ (here $D$ is the laser beam diameter measured at the halfheight of the energy density distribution).

Worth noting, as the focal distance decreases the plasma channel becomes shorter, but its lateral width increases and the luminal emission becomes more intensive. With increasing pulse energy, the integral fluorescence of the plasma filament *B* increases according to a certain power law, and the growth increment depends nonmonotonically on the degree of focusing (Fig. 1b). The maximal grows rate of filament luminescence with the pulse energy is observed for a loosely focused pulse, NA ~ 1.5 - 3·$10^{-3}$, while the minimal rate is characteristic of moderate focusing, NA ~ 4.5·$10^{-3}$.

More clearly, the influence of focusing conditions is demonstrated in Figs. 1(c) and (d), where the integral plasma fluorescence is plotted for two pulse energies and different NA-values. As seen, in the dependencies one can conditionally distinguish the range of numerical apertures showing a marked change in the integral filament fluorescence. This NA-range corresponds to the region of moderate pulse focusing. Within this range, the minimum plasma emission is observed at NA = 4.5·$10^{-3}$ for a carrier of 740 nm, while for shorter-wavelength radiation at 470 nm, the minimum integral fluorescence corresponds to a tighter focusing, NA = 7·$10^{-3}$.

It is worth emphasizing that these features have a certain universality in the sense that the shape of the curves is preserved when the pulse energy is changed. This suggests that in this range of focusing, the predominant influence in the development of filamentation is the geometric compression of laser beam due to the initial curvature of the phase front, rather than self-focusing due to the Kerr effect. This region of NA values can be conditionally associated with the, so-called, transition interval between the quasi-linear and quasi-nonlinear pulse focusing in medium as previously reported in [23, 24]. Within this focusing interval, the partial contributions to the optical wave phase due to the Kerr effect and plasma generation are compared, and the transition from the linear to the nonlinear focusing mode takes place. It was also claimed that this transition region of pulse focused propagation depends weakly on the laser pulse parameters.

To explain the experimental data obtained, we numerically simulate the filamentation of a focused femtosecond laser pulse in air using the methods described in detail in [25]. In numerical experiments, we solve the scalar equation of pulse unidirectional propagation (UPPE) with resolution to the carrier frequency and accounting the optical nonlinearity of the medium [26]. The model of air optical nonlinearity includes the instantaneous and inertial components of the optical Kerr effect, the change in the complex refractive index of air due to photoionization of air molecules and the formation of a plasma of free electrons. The linear part of the UPPE describes the chromatic dispersion of pulse group velocity according to the Cauchy formula and beam diffraction with an intrinsic account of the nonparaxial character of the focused beam propagation. For the calculation the dynamics of free electron volume density $n_e(\mathbf{r},z;t)$ in beam channel, the corresponding plasma rate equation is used which takes into account the combined multiphoton/tunneling ionization of gas components in atmospheric air (nitrogen, oxygen) according to the PPT [27] formulation, avalanche ionization by free electron impacts, as well as the recombination decrease of the electron concentration. The problem is considered in the axisymmetric formulation for the parameters of the initial radiation corresponding to the experimental values.

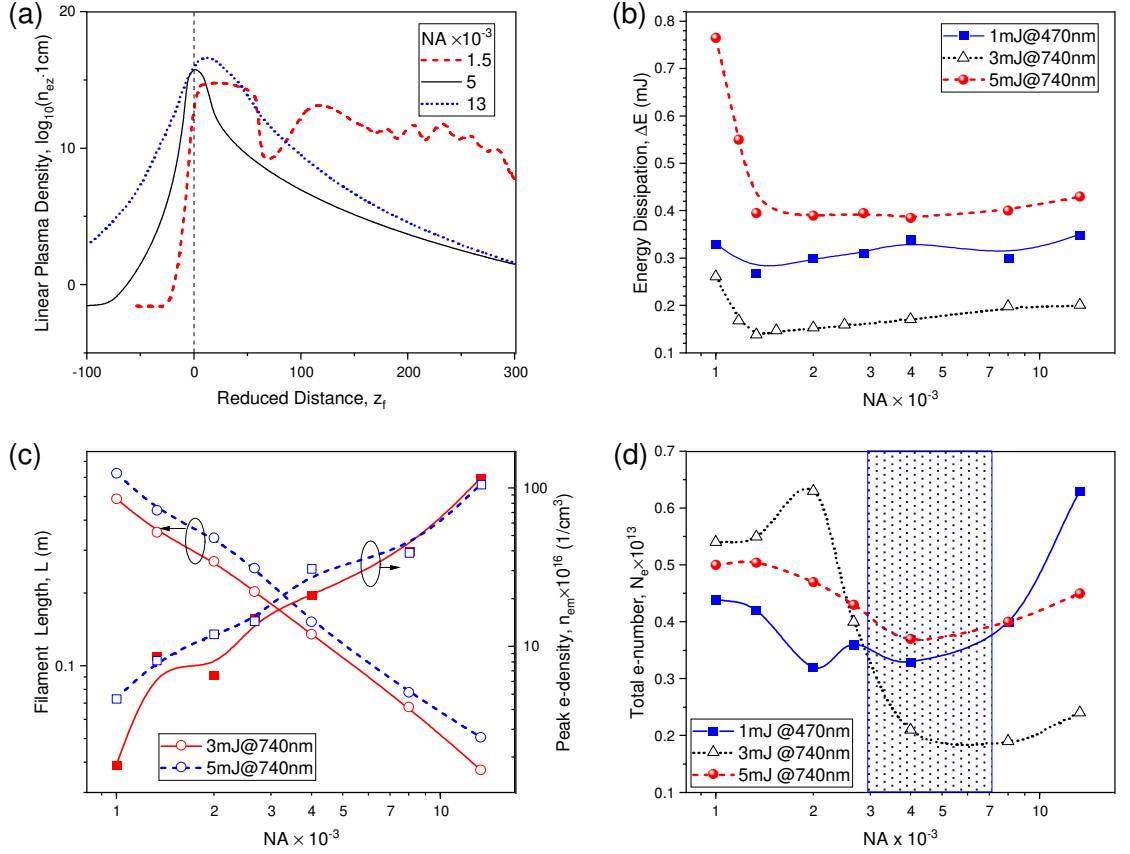

Fig. 2. Simulation of focused pulse filamentation. (a) Linear plasma density $n_{ez}$ as a function of the reduced distance $z_f$ (pulse energy, $E = 3$ mJ). (b) Pulse energy losses $\Delta E$, (c) length $L$ and peak plasma density $n_{em}$, (d) total number of formed plasma electrons $N_e$ as a function of the numerical aperture NA.

Clearly, the main channel of optical pulse energy dissipation during the filamentation is the combined multiphoton/tunnel ionization in the propagation medium and expenditure of pulse energy to the maintaining of filament plasma (inverse Bremsstrahlung absorption). Thus, in the numerical calculations, we mainly analyze the dynamics of several parameters of the filamentation process, namely, the total pulse energy loss $\Delta E$ for the filamentation, linear density of the plasma channel $n_{ez}$, filament length $L$, and the total number of free electrons $N_e$ in the resulting plasma. The integral parameters $n_{ez}$ and $N_e$ are calculated based on the final electron concentration $n_e\left(\mathbf{r}_\perp, z; t \gg t_p\right)$ using the following expressions:

$$n_{ez}(z) = \int_{\mathbf{R}_\perp} n_e(\mathbf{r}_\perp, z) d\mathbf{r}_\perp \qquad (2)$$

$$N_e = \int_{\mathbf{R}} n_e(\mathbf{r}_\perp, z) d\mathbf{r}_\perp dz \qquad (3)$$

The characteristic length of the filament, $L$, is determined at 10 dB drop in electron density relative to the peak value $n_{em}$.

The variation of the linear plasma density $n_{ez}$ along the propagation distance of optical pulse with different energy is shown in Fig. 2(a) for several values of the numerical aperture NA corresponding to conventionally weak (NA = 0.0015), moderate (NA = 0.005), and strong focusing (NA = 0.013). On the horizontal axis in this figure is plotted the reduced distance, $z_f = (z-f)/L_f$, showing the relative spatial remoteness from the geometric focus expressed in focal Rayleigh length, $L_f = 1/k(\text{NA})^2$, where $k$ is the wave number.

It can be seen that moderate beam focusing, in contrast to other extreme values of NA, demonstrates more rapid decrease of $n_{ez}$ values after the focus. At the same time, under conditions of weak focusing, high values of electron density persist on the longest spatial scale, and as a result, the total pulse energy losses $\Delta E$ (Fig. 2b) during the filamentation sharply increase. These losses are associated not only with the increases in the free electron concentration in the filament, but also with the active escape of wave energy through the non-radiation channel to heating of the already formed plasma, which subsequently leads to heating and hydrodynamic expansion of air in the beam channel [13].

For conditionally tight focusing, NA > 0.01, the increase in the pulse energy losses $\Delta E$ is no longer due to the absorption in plasma, but to an increase in the density of laser generated electrons. As shown in Fig. 2(c), the peak values of the electron concentration $n_{em}$ in this range of pulse focusing can reach $10^{18}$ cm$^{-3}$, which is close to the density of neutral air molecules at normal pressure. The filament length $L$ is reduced in this case. Obviously, in general, exactly the opposite tendencies to the growth of $n_{em}$ and reduction of $L$ lead to the appearance of a dip in the dependence of the total number of plasma electrons $N_e$ on the focusing strength, as shown in Fig. 2(d). Indeed, this claim is confirmed if one multiplies the main plasma parameters, ($n_{em} \cdot L$), and analyses the dependence of this product on NA. The obtained dependence will also exhibit a dip in the range of moderate pulse focusing (not shown). Moreover, for the laser wavelength of 470 nm this decrease in energy deposition becomes even more pronounced due to a sharper dependence of the medium ionization on the pulse intensity. It should be noted that the total number of free electrons shown in Fig. 2(d), demonstrates good qualitative and quantitative agreement with the dependences for the integral fluorescence of filament plasma obtained during our experiments (Fig. 1c) that once again confirms the physical consistency of the expression (1).

## 3. Conclusion

In conclusion, we consider the influence of external spatial focusing of a femtosecond laser pulse during its filamentation in air on the amount of optical energy deposited in the propagation medium as a result of laser ionization and plasma generation. The results of our laboratory experiments and theoretical simulations show, for the first time, that in the range of pulse numerical aperture NA, approximately from 0.003 to 0.007, a pronounced decrease (about 50%) of filament energy transferred into air is observed. This focusing range is in close correlation with the discovered earlier transition region between the linear and nonlinear focused propagation of high-power laser pulse [24], which is governed mainly by the action of geometric focusing or Kerr self-compression, respectively. The results obtained can be, for example, a roadmap for the construction the gradient dynamic optical waveguide in air after the passage of a laser filament, or for the optimization the conditions of remote ignition of combustible gases by femtosecond filament plasma.

**Acknowledgement**: YEG acknowledges the Ministry of Science and Higher Education of the Russian Federation (V.E. Zuev Institute of Atmospheric Optics SB RAS).